# Interfacial Dzyaloshinskii-Moriya interaction, surface anisotropy energy, and spin pumping at spin orbit coupled Ir/Co interface

Nam-Hui Kim,[1] Jinyong Jung,[1] Jaehun Cho,[1] Dong-Soo Han,[3] Yuxiang Yin,[3] June-Seo Kim,[3,a)] Henk J. M. Swagten,[3] and Chun-Yeol You[1,2,b)]

[1]Department of Physics, Inha University, Incheon, 22212, South Korea

[2]Department of Emerging Materials Science, DGIST, Daegu, 42988, South Korea

[3]Department of Applied Physics, Center for Nanomaterials, Eindhoven University of Technology, PO Box 513, 5600 MB Eindhoven, The Netherlands

The interfacial Dzyaloshinskii-Moriya interaction (iDMI), surface anisotropy energy, and spin pumping at the Ir/Co interface are experimentally investigated by performing Brillouin light scattering. Contrary to previous reports, we suggest that the sign of the iDMI at the Ir/Co interface is the same as in the case of the Pt/Co interface. We also find that the magnitude of the iDMI energy density is relatively smaller than in the case of the Pt/Co interface, despite the large strong spin-orbit coupling (SOC) of Ir. The saturation magnetization and the perpendicular magnetic anisotropy (PMA) energy are significantly improved due to a strong SOC. Our findings suggest that an SOC in an Ir/Co system behaves in different ways for iDMI and PMA. Finally, we determine the spin pumping effect at the Ir/Co interface, and it increases the Gilbert damping constant from 0.012 to 0.024 for 1.5 nm-thick Co.

---

a) E-mail: spin2mtj@gmail.com

b) E-mail: cyyou@inha.ac.kr

Spin-orbit coupling (SOC) plays a crucial role in various basic magnetic phenomena such as magnetic crystalline anisotropies, magnetostriction effects, magneto-optical Kerr effects, anomalous Hall effects, anisotropic magnetoresistances, and magnetic damping processes.[1] Recently the SOC has been of growing interest due to novel physics and emerging technologies of "spin orbitronics" such as the spin Hall effect,[2,3,4] the Rashba effect,[5,6] the Dzyaloshinskii-Moriya interaction (DMI),[7,8,9] and spin pumping effects.[10,11] Therefore, an understanding the SOC, in particular at the interface between ferromagnetic (FM) material and heavy metals (HM), is of great importance for an investigation of the underlying physics of this exotic phenomena. Among these SOC-related phenomena, an interfacial DMI (iDMI) has drawn recent interest. The iDMI is known to arise at the interface due to inversion symmetry breaking and large SOC in heavy metals (HM). The iDMI manifests itself by forming spiral spin configurations with a preferred chirality. Therefore, it plays an important role in the dynamics of the chiral domain wall (DW)[12,13] and the skyrmion formation.[14,15] Experimental determination of the iDMI energy density is needed. However, general approaches for the iDMI measurements (i.e., asymmetric domain wall motion or bubble expansion) have been difficult to obtain the exact value of the DMI energy density.[16,17,18]

Recently, direct measurements of the iDMI via Brillouin light scattering (BLS) have been demonstrated in Pt/Co/AlO$_x$, Pt/CoFeB/AlO$_x$,[19] and Ta/Pt/Co/AlO$_x$[20] systems. Based on the non-reciprocal spin wave (SW) theorem under the iDMI,[21] BLS can provide a direct value of the iDMI energy density from a measurement of the frequency difference between negative (Stokes) and positive (anti-Stokes) SW frequencies. Therefore, BLS is considered to be a superior approach for the study of iDMI.[22,23,24] Since BLS is able to measure intrinsic material parameters such as the saturation magnetization ($M_S$), the effective magnetic anisotropy energy, and the Gilbert damping constant simultaneously, it is suitable for the measurement of complex SOC-related phenomena accompanying perpendicular magnetic anisotropy (PMA), the enhancement of Gilbert damping by spin pumping, and the enhancement of saturation magnetization by the proximity effect. For example, it is well known that PMA is proportional to the square of SOC strength,[25,26] whereas spin pumping is proportional to the SOC strength.[10,11] Furthermore, the enhancement of proximity-induced magnetization (PIM) is also related to SOC.[27]

In this Letter, we experimentally investigate the interfacial magnetic properties of an Ir/Co($t_{Co}$)/AlO$_x$ system by BLS.[19,20] Iridium, which is a heavy metal, was chosen to compare the SOC at the interface between Cobalt and HM. From systematic BLS measurements, a large value of $M_S$ due to a proximity effect and an enhanced value of $K_S$ due to SOC at the Ir/Co interface were observed. We measured iDMI energy densities from the magnetic field-dependent frequency difference measurements, and found that they are smaller than in the case of the Pt/Co interface. Contrary to previous reports,[28,29,30] the measured sign of the iDMI is the same with Pt/Co/AlO$_x$ systems. Furthermore, a noticeable enhancement of the Gilbert damping constant due to a strong spin pumping effect was observed. This fact implies that the roles of SOC for iDMI, PMA, proximity-induced magnetization, and a spin pumping effect are not the same based on our experimental observations.

The inset in Fig. 1(a) shows the sample geometry. A wedge-shaped Ta(4 nm)/Ir(4 nm)/Co($t_{Co}$)/AlO$_x$(2 nm) sample was deposited on a thermally-oxidized Si wafer. A magnetron sputtering system was used to deposit all the layers. Especially, the Co layer was deposited wedged shape in the range of 1 to 3 nm. In order to break the inversion symmetry, a 2-nm AlO$_x$ capping was used on the top of the Co layer. In our BLS measurements, a p-polarized laser with 300 mW of power and a 532 nm wavelength was used to create and annihilate magnons at the surfaces of the Co layer, which is called the Damon-Eshbach surface mode.[31] More detailed descriptions about BLS measurements are given in the references.[19,32,33]

Figure 1(a) indicates the frequency differences ($\Delta f$) between Stokes and anti-Stokes peak positions in BLS spectra as a function of the Co thickness. Due to the non-reciprocal SW propagation properties with finite iDMI, $\Delta f$ is directly proportional to the iDMI energy density ($D$). The correlation between $\Delta f$ and the $D$ is given by[21]

$$\Delta f = \frac{2\gamma D}{\pi M_s} k_x \qquad (1)$$

where $k_x$ and $D$ are the propagating spin wave (SW) wave vector along the x-direction and the iDMI energy density, respectively. In this study, the in-plane k-vector is fixed at $k_x$ = 0.0167 nm$^{-1}$, corresponding to the back-scattered light by thermal excitation with an angle of incidence of 45$^\circ$.

In Fig. 1(b), the deduced iDMI energy densities based on Eq. (1) are plotted as a function of $t_{Co}^{-1}$. The figure shows inverse proportionality, which is direct evidence that the iDMI at the Ir/Co bilayer is generated at the interface.[19,20] We note that the iDMI energy densities are much smaller than in the case of the Pt/Co interface, as shown in Fig. 1(b).

Next, we considered the magnitude and sign of iDMI for an Ir/Co/AlO$_x$ system. Recently, using the Vienna *ab initio* simulation package (VASP), Yang *et al.*[30] reported that the iDMI of an Ir/Co system is much smaller than those of Pt/Co systems. They also found that the magnitude and sign of the iDMI at Ir/Co is very small and has a direction opposite to the case of the Pt/Co interface, respectively. This is consistent with reported experimental observations of a Pt(111)/Ir($t_{Ir}$)/[Ni/Co]$_N$ and Ta/Pt/Co/Ir($t_{Ir}$)/Pt system.[17,28] Chen *et al.*[28] observed real space images by spin-polarized low-energy electron microscopy (SPLEEM), and they found the transition of DW chirality to be a function of an Ir inserting layer between Pt and [Ni/Co]$_N$. This implies that the iDMI of the interface of Ir/[Ni/Co]$_N$ is opposite to that of Pt/[Ni/Co]$_N$. Hrabec *et al.*[17] used a magneto-optical Kerr effect (MOKE) microscope to observe asymmetric magnetic domain expansion due to the iDMI, and they inserted an Ir layer into the Ta/Pt/Co/Ir($t_{Ir}$)/Pt system. Based on the asymmetric DW velocities, they concluded that the sign of Co/Ir is opposite that of the Co/Pt interfaces.

Contrary to previous theoretical and experimental results,[17,28,30] our experimental data indicates that the sign of iDMI at the Ir/Co/AlO$_x$ interface is the same with the Pt/Co/AlO$_x$, Pt/CoFeB/AlO$_x$, and Ta/Pt/Co/AlO$_x$ systems.[19,20] In order to cross-check our field dependence $\Delta f$ measurements (see Fig. 1(b)), SW propagation angle dependence measurements were performed. The relation between $\Delta f$ and $\theta_\alpha$, the angle between the direction of the propagating SWs and the direction of the applied field, is given by[34]

$$\Delta f(\theta_\alpha) = \Delta f_0 \sin \theta_\alpha, \qquad (2)$$

where $\Delta f_0$ is the maximum $\Delta f$ when $\theta_\alpha$ equals $\pm\pi/2$; this is a typical measurement geometry for our magnetic field and SW wave vector dependence measurements. The angle $\theta_\alpha$-dependent $\Delta f$ measurements from -90° to +90° with a 22.5° interval for the Ir/Co ($t_{Co}$ = 1.75 nm and 2.0 nm) and Pt/Co ($t_{Co}$ = 1.8 nm) cases are shown in Fig. 2. All sinusoidal fitting curves are well matched with the measured $\Delta f$ as a function of $\theta_\alpha$. The green area showing that Δ$f$ is less than about 0.3 GHz indicates the limit of our BLS experiments. Consequently, we suggest that the signs of iDMI are the same at Ir/Co and Pt/Co interfaces.

Currently, the physical origin of the sign of iDMI at Ir/Co/AlO$_x$ is not fully understood. However, there are several possible reasons for the opposite sign reported in previous experimental results.[17, 28] Our measurements were conducted on Ir(4 nm)/Co/AlO$_x$, whereas the previous results had different sample stacks; i.e., Ta/Pt/Co/Ir/Pt[17] and Pt(111)/Ir/[Ni/Co]$_N$.[28] For example, Hrabec *et al.*[17] inserted an Ir layer on the top of a Co layer and covered it with Pt capping. Chen *et al.* investigated the Ir/Ni and Co/Ir interface,[28,35] not the Ir/Co interface. Based on our experimental results, we suggest that the iDMI energy density is quite sensitive to the details of the multilayer structures. Moreover, the iDMI is not only determined by the nearest interfaces, it may have long range characteristics, as predicted by Fert and Levy.[36]

We investigate other important and SOC-related quantities in a Ta/Ir/Co/AlO$_x$ system by using BLS measurements. We ignored the effect of iDMI on the SW excitation frequency, and took the average of the Stokes and anti-Stokes SW frequencies without the iDMI contribution. This can be expressed as[37]

$$f_{\rm SW} = \frac{\gamma}{2\pi} \sqrt{H_{\rm ex}\left(H_{\rm ex} - \frac{2K_{\rm eff}}{\mu_0 M_{\rm S}}\right)} \qquad (3)$$

Here, we ignored the exchange energy and the bulk PMA contribution. $K_{\rm eff} = \frac{2K_{\rm S}}{t_{\rm Co}} - \frac{1}{2}\mu_0 M_{\rm S}^2$, $K_{\rm S}$ is the surface anisotropy energy, $\gamma$ is the gyromagnetic ratio, and $H_{\rm ex}$ is the applied magnetic field. We can extract the effective anisotropy energy $K_{\rm eff}$ from the external field-dependent $f_{\rm sw}$ using Eq. (3), and $K_{\rm eff} \times t_{\rm Co}$ versus $t_{\rm Co}$ is shown in Fig. 3. From the slope and $y$-intercept, the saturation magnetization $M_{\rm S}$ (=1.68±0.02×10$^6$ A/m) and the surface anisotropy energy $K_{\rm S}$ (=1.36±0.01 mJ/m$^2$) were determined, respectively. For $t_{\rm Co}$ > 1.5 nm, the effective anisotropy becomes negative such that the easy axis of the sample is in-plane. Surprisingly, the obtained values of $K_{\rm S}$ and $M_{\rm S}$ at the Ir/Co interface are significantly enhanced in comparison with the case of the Pt/Co ($M_{\rm S}$= 1.42±0.02×10$^6$ A/m) interface.[20] Especially, the measured $M_{\rm S}$ (=1.68×10$^6$ A/m) is 20% greater than the bulk magnetization of Co ($M_{\rm S}$ = 1.4×10$^6$ A/m). Experimental evidence that a large PIM exists at the Ir/Co interface has recently been reported.[27] The reported PIM in an Ir/Co/Ni/Co system is 19%, which is quite similar to our observed value.

The $K_S$ value (1.36 mJ/m$^2$) of the Ir/Co system is noticeably enhanced compared to the values for Pt/Co/AlO$_x$ ($K_S$ = 0.54 mJ/m$^2$) and Ta/Pt/Co/AlO$_x$ ($K_S$ = 1.1 mJ/m$^2$) in our previous reports.[19,20] On a theoretical basis, it has been reported that Ir monolayer capping induces the strongest surface PMA of an Fe(001) layer.[38] They found that the PIM and the PMA of Ir is even larger than that of Pt. This gives a clue regarding the observed enhancements of the values of $M_S$ and $K_S$ in our Ir/Co/AlO$_x$ system. Regardless, their study is about 5$d$ transition metals with Fe, and not Co. Furthermore, Broeder *et al.* reported that $K_S$ for Ir/Co (~ 0.8 mJ/m$^2$) is larger than that of Pt/Co (0.5 ~ 0.58 mJ/m$^2$).[39]

Another important effect of a strong SOC is the enhancement of Gilbert damping due to a strong spin pumping effect.[10,11] The precession of spins in a ferromagnetic layer induces a spin current in the adjacent layer and a loss of angular momentum, and causes additional damping. The amount of spin pumping is closely related to the SOC through the spin flip relaxation time and the interface mixing conductance. As a result, spin pumping is an important path for the magnetic damping of HM/FM structures. It has been reported that spin pumping can be suppressed by interface engineering, or by introducing a nano-oxide layer between HM and FM by using vector-network analyzer ferromagnetic resonance (VNA-FMR).[11]

We obtain a full-width at half maximum (FWHM) from each resonance frequency spectrum from BLS SW spectra, similar to the VNA-FMR experiment. To extract the Gilbert damping constant, we applied a modified equation which used from FMR system as the condition of the applied large in-plain magnetic field in a PMA system. The FWHM ($\Delta f_{\text{res}}$) has the following relation with the Gilbert damping constant $\alpha$:

$$\Delta f_{\text{res}} = \frac{2\alpha\mu_0\gamma}{\pi} H_{\text{ex}} + \Delta f_{\text{res}}^{\text{extrinsic}} \qquad (4)$$

where $\Delta f_{\text{res}}^{\text{extrinsic}}$ is the additional linewidth at the resonance frequency by an extrinsic source. Figure 4(a) clearly shows the linear relations between linewidths ($\Delta f_{\text{res}}$) and the applied magnetic fields ($H_{\text{ex}}$) for $t_{\text{Co}}$ = 1.5 and 3.0 nm.

We examined another possible mechanism of FWHM broadening: the two-magnon scattering (TMS) process. It is well known that TMS occurs when $\theta_\alpha(\vec{k}_\parallel)$, the angle between the SW propagation direction and the in-plane field (or magnetization direction), is smaller than the critical angle $\theta_c$.[40] Here, the critical angle $\theta_c = \sin^{-1}\left[\frac{H_{ex}}{B_0+H_S}\right]^{\frac{1}{2}} = 41.2°$ when the in-plane external field $H_{ex}$ is 0.2 T. $B_0 = H_{ex}+4\pi M_S$; $H_S = 2K_S/M_s d$; and $M_S$, $K_S$, and $d$ are the saturation magnetization, the surface anisotropy, and the thickness of the cobalt, respectively. The necessary condition of TMS is $\theta_\alpha(k_\parallel) < \theta_c$ for $k_\parallel > 0$, which corresponds to the anti-Stokes case in our experimental geometry. Since we extracted the FHWM from the anti-Stokes peaks and $\theta_\alpha(\vec{k}_\parallel)=90°$ configuration, there is no contribution from TMS. Furthermore, if TMS contributes to the linewidth, we require a non-zero $\Delta f$ for the $\theta_\alpha(\vec{k}_\parallel)=0$ case. However, we already show a negligible $\Delta f$ for $\theta_\alpha(\vec{k}_\parallel)=0$ in Fig. 2. Therefore, we can exclude the TMS contribution in the observed FWHM broadening.

From the slopes of the linear fittings, the magnetic damping constants $\alpha$ for each Co thickness were deduced and are shown in Fig. 4(b). Normally, the energy dissipation in a magnetic system depends on the imaginary part of the susceptibility of the magnetic system. It has been claimed that the imaginary part of the eigenfrequency is modified by a factor of $(1 + f_{DM}(k)/f_0(k))$ due to the iDMI.[23] Here, $f_{DM}$ and $f_0$ are the resonance frequency with iDMI and without iDMI, respectively. For general cases ($\frac{f_{DM}}{f_0} \ll 1$), the enhancement of damping due to the iDMI is not significant. However, since the observed damping constant enhancement in our experiment is about double at $t_{Co} = 1.2$ nm compared with $\alpha_{Bulk}$, the enhancement due to the iDMI is about 10%. Therefore, we can rule out a contribution due to iDMI in the enhancement of the damping constant.

Figure 4(b) shows that the $t_{Co}^{-1}$ dependence of the damping constant is mainly due to the spin pumping effect. Consequently, due to spin pumping at the Ir/Co interface, an enhancement of the damping constant ($\alpha = 0.02$) is observed at $t_{Co} = 1.5$ nm. With increasing $t_{Co}$, the measured $\alpha$ decreases as shown in Fig. 4(b), since spin pumping is a kind of interface effect. Therefore, the spin pumping is being smeared away when the thickness of the FM layer increases. In Fig. 4(b), the measured $\alpha$ versus $t_{Co}^{-1}$ and a linear dependency with a finite *y*-intercept is seen. The physical meaning of the damping at $t_{Co}^{-1} = 0$ ($t_{Co}\to\infty$) is the damping constant ($\alpha_\text{bulk}$) of bulk cobalt. In these measurements, we determined that $\alpha_\text{bulk} \sim 0.012$, which is in good agreement with the magnetic damping constant for bulk Co ($\alpha_\text{Co}^\text{bulk} = 0.011$).[41]

In conclusion, we used BLS to observe SOC-related physical quantities such as the interfacial Dzyaloshinskii-Moriya interaction, surface magnetic anisotropy, and the magnetic damping constant accompanying the spin pumping effect at the Ir/Co interface. From systematic BLS measurements, we suggest that the measured iDMI energy density is relatively smaller than in the case of the Pt/Co interface. On the other hand, the saturation magnetization and the surface magnetic anisotropy are significantly improved due to a higher proximity-induced magnetization. From this result, we believe that the iDMI and PMA behave in different ways at the Ir/Co interface. Based on the results in previous reports, the sign of the iDMI at the Ir/Co interface is the same as that of the Pt/Co interface.

**Acknowledgement**

This work is supported by the Research Programme of the Foundation for Fundamental Research on Matter (FOM), which is part of the Netherlands Organisation for Scientific Research (NWO) and the National Research Foundation of Korea (Grant nos. 2015M3D1A1070467, 2013R1A1A2011936, and 2015M2A2A6021171).

# Figure Captions

Fig. 1. (a) iDMI induced spin-wave frequency differences ($\Delta f$) as function of $t_{Co}$. (b) iDMI energy density as a function of $t_{Co}^{-1}$ for the external magnetic field dependence ($D_H$) in Ir/Co (black squares), Pt/Co/AlO$_x$ (blue circles), and Ta/Pt/Co/AlO$_x$ (red triangles), respectively. The iDMI energy density of the Ir/Co interface has a maximum value (0.7 mJ/m$^2$) that is much smaller than that of Pt/Co/AlO$_x$ (1.3 mJ/m$^2$) and Ta/Pt/Co/AlO$_x$ (1.7 mJ/m$^2$).

Fig. 2. The frequency differences ($\Delta f$) between Stokes and anti-Stokes from $\theta_\alpha$= -90° to +90° in Ta/Ir/Co/AlO$_x$ ($t_{Co}$=1.75 nm and 2.0 nm) and Ta/Pt/Co/AlO$_x$ ($t_{Co}$=1.8 nm). The measured $\Delta f$ as a function of the $\theta_\alpha$. The solid line is the fitting curve from Eq. (2).

Fig. 3. $K_{eff} \times t_{Co}$ versus $t_{Co}$ plot with a linear fitting. $K_S$ and $M_S$ were extracted from the slope and y-intercept. For $t_{Co}$ > 1.54 nm, the effective uniaxial anisotropy becomes negative, which means the direction of the easy axis changes perpendicular to the in-plane.

Fig. 4. (a) The linewidths as a function of $H_{ex}$ for Ta (4 nm)/Ir (4 nm)/Co($t_{Co}$)/AlO$_x$ (2 nm). The black and red open symbols are experimental values from each of spectrum, and the solid lines are results of linear fitting ($t_{Co}$ = 1.5, 3.0 nm). The error bars were obtained from Lorentzian fitting of SW peaks. (b) The Gilbert damping parameters as function of inverse Co thickness. The extrapolated $\alpha$ for infinitely-thick Co is approximately 0.012, which is well-matched with the damping constant of bulk cobalt ($\alpha_{Co}^{Bulk}$= 0.011).

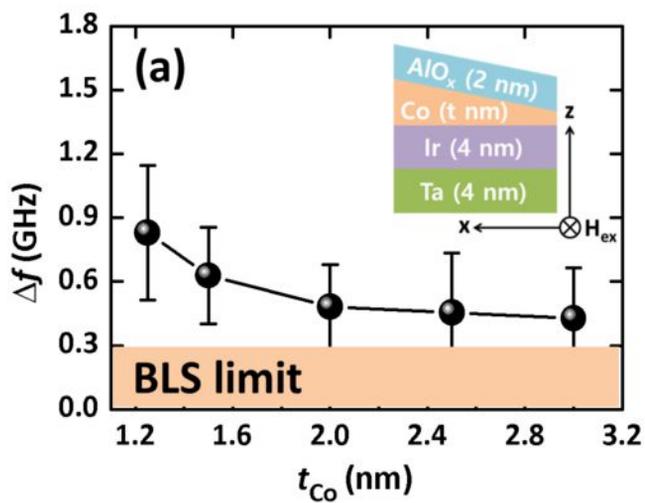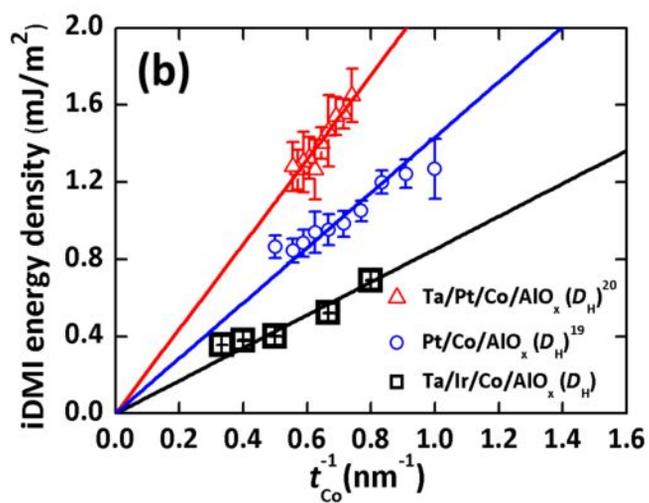

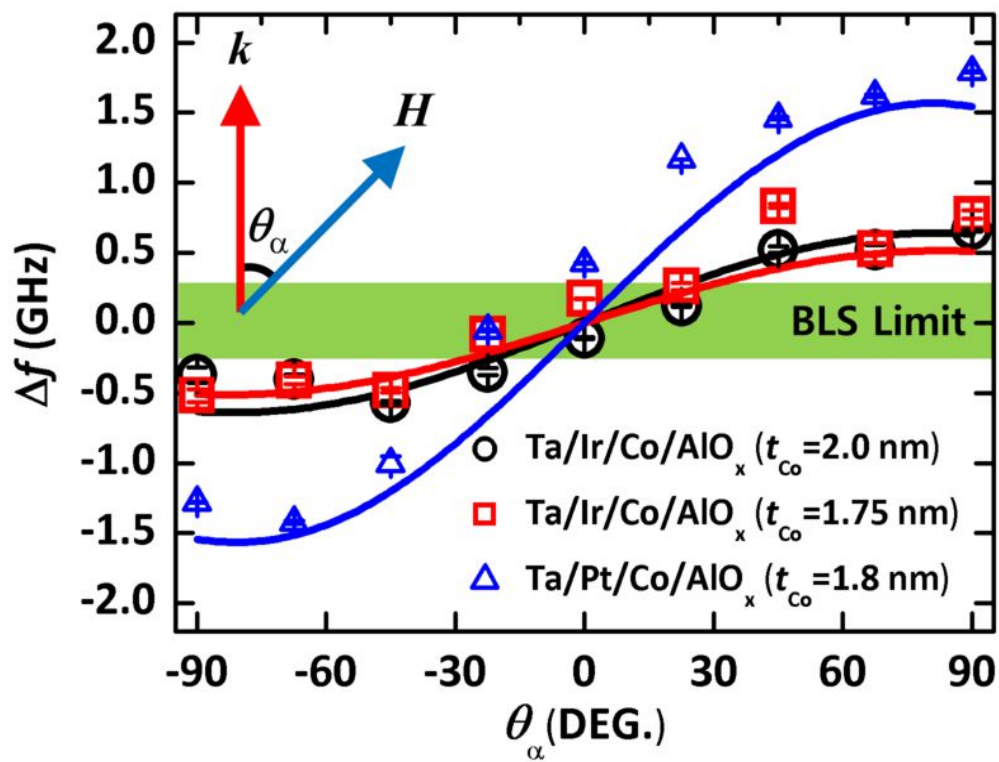

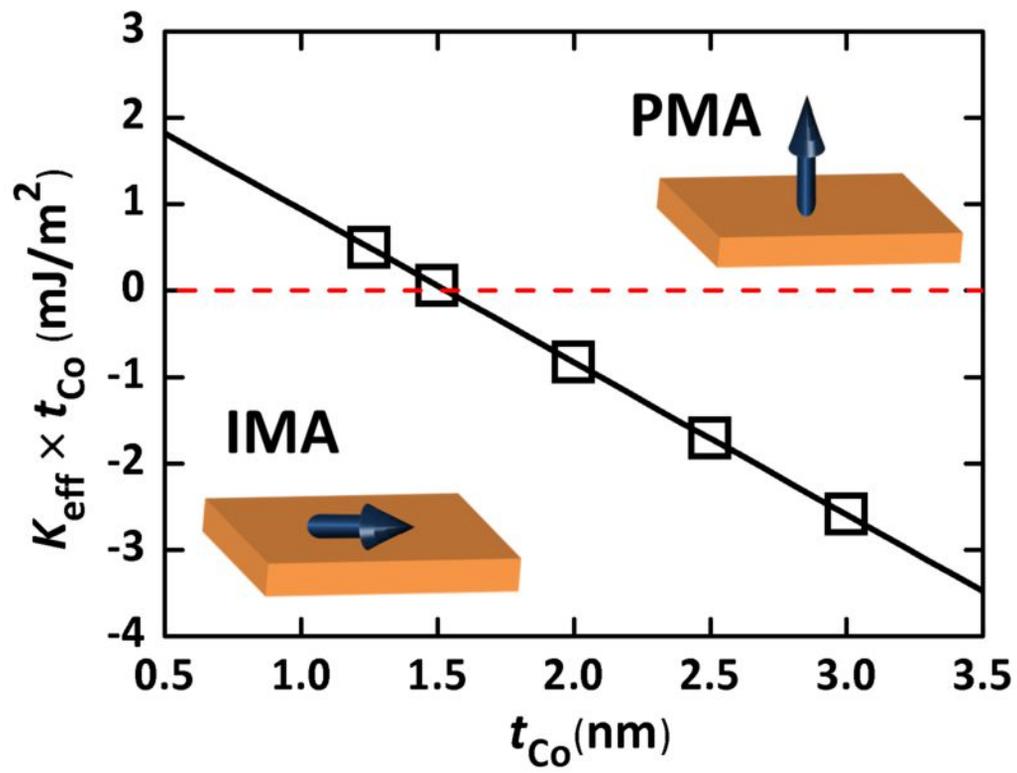

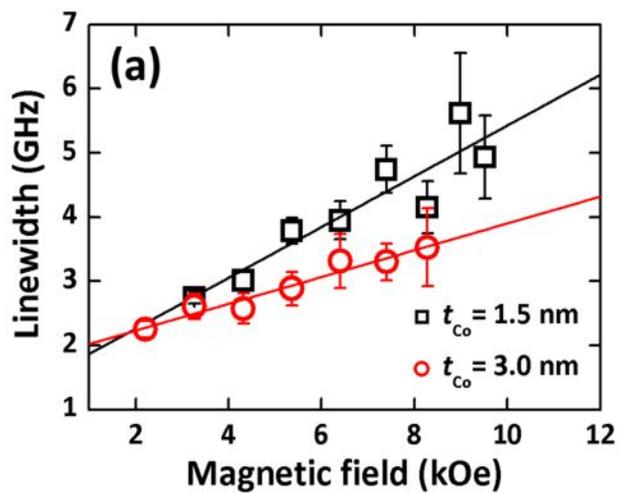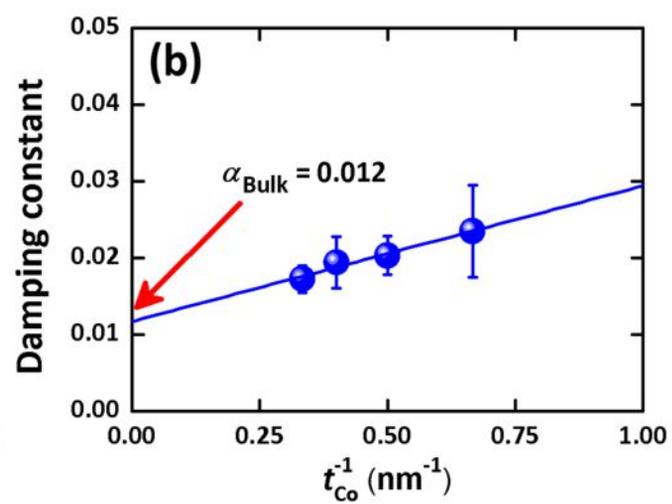